# Superconducting materials: Challenges and opportunities for large-scale applications

Chao Yao[1,2] and Yanwei Ma[1,2,*]

*Correspondence: ywma@mail.iee.ac.cn

[1]Key Laboratory of Applied Superconductivity, Institute of Electrical Engineering, Chinese Academy of Sciences, Beijing 100190, China

[2]University of Chinese Academy of Sciences, Beijing 100049, China

## Summary

Superconducting materials hold great potential to bring radical changes for electric power and high-field magnet technology , enabling high-efficiency electric power generation, high-capacity lossless electric power transmission, small light-weighted electrical equipment, high-speed maglev transportation, ultra-strong magnetic field generation for high-resolution magnetic resonance imaging (MRI) systems, nuclear magnetic resonance (NMR) systems, the future advanced high energy particle accelerators, nuclear fusion reactors and so on. The performance, economy and operating parameters (temperatures and magnetic fields) of these applications strongly depend on the electromagnetic and mechanical properties, as well as manufacturing and material cost of superconductors. This perspective examines the basic properties relevant to practical applications and key issues of wire fabrication for practical superconducting materials, and describes their challenges and current state in practical applications. Finally, future perspectives for their opportunities and development in the applications of superconducting power and magnetic technologies are given.

# Introduction

For many metals and compounds, when cooled to a sufficiently low temperature, their resistivity suddenly drops to zero. This phenomenon known as superconductivity, which is first observed by Dutch physicist Heike Kamerlingh Onnes. In 1908, Kamerlingh Onnes succeeded in liquefying helium at a temperature of 4.2 K, and then in 1911, when he measured the low-temperature resistivity of metals, he found the superconductivity of mercury at 4.2 K [1]. After discovering the zero resistance of the superconductor, in 1933 German physicists W. Meissner and R. Ochsenfeld found that if a superconductor was cooled below the transition temperature $T_c$ in the magnetic field, the magnetic field would be completely ejected from the superconductor. This phenomenon is called the Meissner effect [2], which is another essential characteristic of superconductivity. After that, researchers observed superconductivity in many other substances, and some of them have higher superconducting transition temperatures. At the same time, due to the exotic nature of superconductors, people have also carried out extensive research for their practical applications. The zero resistance and high current density have a profound impact on electrical power transmission and also enable much smaller and more powerful magnets for motors, generators, energy storage, medical equipment, industrial separations and scientific research, while the magnetic field exclusion provides a mechanism for superconducting magnetic levitation, as shown in Figure 1. Owning to the different operating temperature ranges and required magnetic fields, and also the cooling approaches and material properties, currently the industrial applications of superconductors can be categorized into applications such as power cables, fault current limiters, transformers and induction heaters at 65-77 K with liquid nitrogen as coolant and field <1 T, applications such as motors, generators, maglev, energy storage devices, magnetic resonance imaging (MRI) systems and magnetic separations at temperatures below 50 K and fields above 1 T, and high-field magnets (>10 T) for fusion, accelerator, high-field MRI, nuclear magnetic resonance (NMR), and scientific research at low-temperature region (usually at 4.2 K using liquid helium as coolant).

Since the discovery of superconductivity in mercury, lots of superconducting materials were found. According to their constituents and structures, superconducting materials can be divided into several categories: 1) Metallic materials [3], which include pure metals (mercury, lead, niobium etc.), alloys (such as Nb-Ti and Nb-Ge) and intermetallic compounds (such as $Nb_3Sn$, $Nb_3Al$ and $MgB_2$); 2) ceramic compounds, including Chevrel compounds (e.g., $PbMo_6S_8$ and $SnMo_6S_8$) [4], copper-based oxides [3], ruthenium-based oxides [5], and iron-based pnictides and chalcogenides [6]; 3) organic materials (e.g., $K_3C_{60}$, $Cs_3C_{60}$ and $Ba_4C_{60}$) [7-9]; 4) semiconductor, semi-metal and insulators (e.g., SiC, diamond and graphite) [10-12]. In the early research for superconductors, it was found that the superconducting state is not only related to the temperature, but also to the external magnetic field and the current in the superconductor. When the magnetic field applied to the superconductor is larger than a certain critical value $H_c$, the superconducting state will be destroyed. When the current passing through a superconductor is higher than a critical current $I_c$, the superconducting state will also be destroyed, even if the external magnetic field is not applied. Therefore, the applicable range of superconducting materials is primarily limited by these three parameters. So far, though thousands of superconductors have been discovered, the ones with practical value are limited to Nb-Ti, $Nb_3Sn$, copper-based oxide superconductors, $MgB_2$ and iron-based superconductors, as summarized in Table 1.

During the years from 1911 to 1932, low-temperature superconductors (LTS) such as lead, tin, niobium and other metal were found to be superconductors, and among them niobium has the highest $T_c$ of 9.2 K. In the following decades, many alloys, and carbon- and nitrogen-based compounds with superconductivity were discovered. Among these superconducting alloys and intermetallic compounds, Nb-Ti and $Nb_3Sn$ reported in 1961 and 1954, respectively, are the most promising ones for practical applications, with a $T_c$ of 9.5 K and 18.1 K, respectively. At 4.2 K, Nb-Ti and $Nb_3Sn$ have an upper critical field of 11 and 25 T, respectively. Both of them have current densities over $10^5$ A/cm², which are about 2 orders of magnitude higher than that of copper conductors. Therefore, Nb-Ti and $Nb_3Sn$ enabled the construction of superconducting magnets that can generate much higher magnetic fields than conventional resistive magnets.

In 1986, J. Bodnorz and K. Muller discovered LaBaCuO superconductors with a $T_c$ of 35 K, which open the gate of searching for high-temperature superconductors (HTS) [13], as shown in Figure 2. In 1987, the $T_c$ in this system was rapidly increased above the liquid nitrogen temperature (77 K) for the first time because of the discovery of $YBa_2Cu_3O_x$ (YBCO or REBCO, RE = Rare Earth) superconductors with $T_c$s up to 93 K [14, 15]. Then bismuth-based cuprate superconductors (BSCCO) including $Bi_2Sr_2CaCu_2O_8$ (Bi-2212) and $Bi_2Sr_2Ca_2Cu_3O_{10}$ (Bi-2223) with $T_c$s up to 110 K were discovered [16]. As presented in Figure 3, regarding the operating temperatures and magnetic fields, Bi-2223 and REBCO can carry large supercurrents up to 30-50 K, in field and at 77 K in self-field, so they are promising not only for high field magnets operated in low or moderate temperature region but also for electro-technical applications with much cheaper liquid nitrogen as coolant. On the other hand, though Bi-2212 can be used only at low temperature (<20 K), it has its own advantages for high-field applications. In 2001, the superconductivity at 39 K in $MgB_2$ was discovered by Akimitsu group in Aoyama-Gakuin University [18]. This transition temperature is the highest so far for the bulk binary intermetallic compounds. $MgB_2$ is promising for applications at around 20 K that can be easily achieved by liquid hydrogen or cryocoolers, and is considered to replace traditional low-temperature superconductors such as Nb-Ti and $Nb_3Sn$ used in liquid helium. Moreover, the abundant raw materials and light weight of $MgB_2$ also make it attractive for large-scale practical applications. In 2008，the discovery of iron-based superconductors [19] by Hosono group in the Tokyo Institute of Technology marked the coming of the 'iron age' of high-$T_c$ superconductivity after the 'copper age' marked by cuprate superconductors. According to different chemical compositions and crystal structures, iron-based superconductors can be categorized into several types, such as '1111' type (e.g., $LaFeAsO_{1-x}F_x$ and $SmFeAsO_{1-x}F_x$), '122' type (e.g., $Ba_{1-x}K_xFe_2As_2$ and $Sr_{1-x}K_xFe_2As_2$), '111' type (e.g., LiFeAs) and '11' type (e.g., FeSe, $FeSe_{1-x}Te_x$). Similar to cuprate superconductors, iron-based superconductors have layered crystal structure and small coherence lengths, showing high upper critical fields and electromagnetic anisotropy. Though the $T_c$s of iron-based superconductors (up to 38 K for '122' system and 56 K for '1111' system) are not as high as that of cuprate superconductors, their anisotropy of is remarkably lower, especially for '122' and '11' systems. The high upper critical fields and low anisotropy make iron-based superconductors being quite attractive for high-field applications that can work at liquid helium temperature and also in moderate temperature range around 20 K.

Very recently, room-temperature superconductivity, which had always been a dream of researchers over the past 100 years, was reported in a carbonaceous sulfur hydride with a critical temperature up to 287.7 K (~15 degrees Celsius) under an extremely high pressure of 267 GPa [20], as shown in Figure 2. However, there is still doubt that whether this result belongs to a novel kind of superconductivity that different from the standard conventional and unconventional superconductivity, or was misinterpreted as superconductivity [21]. From the viewpoint of practical applications, the operation at such high pressure is much more difficult than that at low temperatures. Nevertheless, this result will encourage the exploration for practical room-temperature superconducting materials in the future.

## Material challenges and current state for practical applications

For high current and/or high magnetic field applications, superconductors must be made into composite wires for cabling or coil winding. Except for large current carrying capacity (indexed by critical current density $J_c$, for which $10^5$ A/cm$^2$ at the operating temperature and magnetic field is widely accepted as the threshold for practical applications), superconducting wires are required to have high mechanical strength to withstand the electromagnetic and thermal stress during operation, fine superconducting filaments in metal matrix to protect against flux jumps and thermal quenching and sufficient length of hundred/ thousand meters for large-scale use [22]. As shown in Figure 4, nowadays low-temperature superconductors such as Ni-Ti and Nb$_3$Sn have been fully commercialized, and widely applied to electric power industry. The 1$^{st}$ Generation HTS (Bi-2223 and Bi-2212) are on the stage of medium-term commercialization, with which many commercial demonstration projects have been completed. In the recent years, the 2nd Generation HTS (REBCO) has undergone rapid development, and entered the stage of early commercialization. $MgB_2$ wires produced with powder-in-tube (PIT) method are also on the early stage of commercialization, while the ones made with internal Mg diffusion method are still being researched in labs. For iron-based superconductors, which are still under laboratory research and development, rapid progress has been made for 122-type wires, and so far some attempts for experimental demonstrations have been made.

**1. Nb-Ti and Nb$_3$Sn superconductors**

Owing to its nice ductility, Nb-Ti alloy can be directly deformed into long wires from monofilament billets with copper matrix and multifilament billet made by assembling tens of thousands of monofilament wires. Soon after its discovery, Nb-Ti superconducting wires was fully industrialized in 1968 and widely used in practical applications. It should be noted that currently Nb-Ti alloy is still the cheapest practical superconducting material for applications in the liquid helium temperature region, because the raw materials and manufacturing costs are much lower than other superconducting materials. On the other hand, due to the brittleness of materials, $Nb_3Sn$ cannot be directly deformed in to wires as done for Nb-Ti alloy. Several types of processes can be used for wire fabrication, including the bronze process (started with Nb alloy rods clad in ductile Cu-Sn bronze and then assembled inside another Cu-Sn tube to obtain multifilament structure), internal Sn process (started with placing a Sn core in the center of Nb-filament embedded Cu matrix) and PIT process (started with packing Sn-rich Nb-Sn powders such as $Nb_6Sn_5$ and $NbSn_2$ in Nb tubes). By heat treatment for cold-worked wires, the $Nb_3Sn$ superconducting phase can be obtained by reaction between Sn and Nb. The key to improve the non-Cu $J_c$ of $Nb_3Sn$ wires is the enhancement of their pinning capacity. Since the grain boundary pinning in $Nb_3Sn$ is neither as efficient nor as dense as α-Ti pinning in Nb-Ti alloy, grain refinement and artificial pinning centers are effective for flux pinning [30]. Due to the increased complexity of wire manufacture, the commercialization of $Nb_3Sn$ superconducting wires was realized after 1970, with a relatively higher price than that of Nb-Ti superconductors.

Since the 1960s, Nb-Ti and Nb$_3$Sn superconductors have greatly promoted the development of superconducting magnets, and thus stimulate the industry for superconducting materials and technologies. Nb-Ti superconductors are usually used to manufacture superconducting magnets that generate magnetic field up to 9 T at 4.2 K or 11 T at 1.8 K. At present, Nb-Ti superconducting wires are mainly used in the construction of MRI systems, superconducting magnets for laboratories, magnetic levitation trains and so on, with a consumption of about several thousands of tons each year. For Nb$_3$Sn superconductors, their niche in the market is high-field applications beyond the capability of Nb-Ti and in the range up to ~23 T. The important application areas of Nb$_3$Sn superconductors include MRI systems, NMR devices, particle accelerators, tokamak fusion devices, and high-field magnets for scientific research. Taking the fusion applications for example, from 2008 to 2015 more than 500 tons of Nb$_3$Sn wires were procured for the international thermonuclear experimental reactor (ITER) project, which led to a ten-fold boosted global Nb$_3$Sn production capability. Now the cutting-edge Nb$_3$Sn wires are the internal-tin restacked-rod-process (RRP®) wires with the highest non-Cu $J_c$ of 1.6-1.7×10$^5$ A/cm$^2$ (4.2 K, 15 T), which is getting close to the requirement (1.5×10$^5$ A/cm$^2$ at 4.2 K, 16 T) of more than 5000 dipole magnets for the Future Circular Collider [31]. For further improving the $J_c$ performance, Hf and Zr addition to form ternary alloys Nb-Ta-Hf and Nb-Ta-Zr and nano inclusions introduced with internal oxidation technique to reduce the grain size have shown great potential [32, 33]. Industrial fabrication for Nb-Ta-Hf alloy with average grain size less than 50 μm has been achieved by ATI metals [30].

**2. Copper-oxide superconductors**

The copper-oxide superconducting materials have high $T_c$ above the liquid nitrogen temperature (77 K) and even liquefied natural gas (LNG) temperature (113 K). Due to the extremely rich abundance of nitrogen, the cost of refrigeration with liquid nitrogen is much lower compared with liquid helium, making it possible for the large-scale industrial applications based on superconducting technology. Though the $T_c$ of the three cuprate superconducting compounds Bi-2223, Bi-2212 and REBCO are much higher than that of Nb-Ti and Nb$_3$Sn, they are much more difficult to be processed into wires and tapes due to their ceramic brittleness. The plate-like crystals of cuprate compounds are formed during a high temperature heat treatment, during which the oxygen content in these ceramic compounds should be controlled to obtain an optimal superconducting phase. In addition, there is weak-linking effect between their grains with large grain boundary angles, which is also not beneficial to the current carrying ability and can be reduced by grain texture. For Bi-2223 superconductors, aligned grains with misalignment at the c-axis <15° is essential for high inter-grain critical currents. Since Bi-2223 is a metastable phase and will decompose during melting, the grain orientation can be obtained by mechanical deformation (rolling) and heat treatment process. Therefore, based on the PIT process, high performance Bi-2223 strands are manufactured with an appearance of tape. For Bi-2212 superconductors, PIT wires or tapes can be fabricated because well connected and textured grains can be obtained from the liquid phase at a temperature lower than the melting point of silver sheath. The PIT technique is not applicable to REBCO since it carries large critical current only in highly biaxially textured tape samples with grain misalignment <5°, which is hard to achieve by the PIT technique. It only took a few years from the discovery to the commercialization of the first Bi-based wires and tapes because of the achievement of mechanical deformation induced grain texture. In contrast, REBCO coated conductor tapes (a few meters long) were first manufactured almost 20 years after the discovery of the compound and after about 10 years of intense R&D effort in the US, Europe, and Japan. The 1990s were a highly active period in the development of Bi-based wires and tapes, while most of the breakthroughs in REBCO coated conductors were achieved during the 2000s [34].

2.1 BSCCO

For the fabrication of BSCCO wires and tapes, the PIT method introduced above for $Nb_3Sn$ is used for both Bi-2223 and Bi-2212 compounds. Starting powders, such as oxides and carbonates, are mixed and calcined to obtain precursors that are packed into a metal sheath, which is then mechanically deformed into wires and tapes. Ag or Ag alloy is required as sheath material instead of Cu and other metals for BSCCO wires and tapes, because they are chemically compatible (i.e., Ag hardly reacts with the precursor) and transparent to oxygen (i.e., Ag can transmit the oxygen released from or absorbed into oxide superconductors during heat treatment). Multifilamentary wires can be fabricated by further restacking monofilamentary wires in an Ag sheath and then repeating the cold deformation process. To strengthen the coupling between BSCCO grains, it is necessary to make the randomly oriented grains aligned. Since BSCCO oxides are crystallized with plate-like appearance due to their high anisotropy, the grain orientation is relatively easy. However, the approaches of grain orientation is different for Bi-2212 and Bi-2223. For Bi-2212 wires, partial melting followed by gradual cooling is adopted, while Bi-2223 tapes are processed by rolling induce texture. In both cases, microstructures with uniaxial (c-axis) orientation are obtained. In 2005 Sumitomo [35] developed a controlled over-pressure process which decreased the porosity and improved grain connectivity; as a consequence, the production yield was increased and $I_c$ exceeded 200 A at 77 K, self-field. While for Bi-2212 wires, the transport $J_c$ did not change much till 2011, when it was found that the porosity in the ceramic (due to the formation of gas bubbles during heat treatment), rather than grain misalignment, was the main obstacle to transport supercurrent [36]. With a high pressure heat treatment the transport $J_c$ of Bi-2212 wires can be increased by two times to $4\times10^5$ $A/cm^2$ (4.2 K, 15 T) [37]. Moreover, homogeneous precursor powder with low impurity can helps to reduce the porosity, thus further raising the transport $J_c$ to $6.6\times10^5$ $A/cm^2$ (4.2 K, 15 T) [25].

For Bi-2223 commercial tapes, companies including American Superconductor (AMSC), Sumitomo (Japan), Bruker (Germany) and Innova Superconductor Technology (China) were able to produce kilometer-class long tapes. Bi-2223 tapes have been used in many demonstration projects for power cables, motors, generators, transformers, and fault current limiters across the world [38]. In 2012, the world's first

HTS power substation, which was developed by the Institute of Electrical Engineering, Chinese Academy of Sciences (IEECAS), was put into operation officially in the power grid in Baiyin city, Gansu province, China. The substation, which integrates a superconducting magnetic energy storage device, a superconducting fault current limiter, a superconducting transformer and an AC superconducting transmission cable, can enhance the stability and reliability of the grid, improve the power quality and decrease the system losses [39]. With laminated mechanical reinforcement technique by Sumitomo for the weak Ag sheathed Bi-2223 tapes, the mechanical strength of the tapes is significantly enhanced, making them an alternative for high-field applications. A 24.6 T cryogen free magnet with a Bi-2223 insert and an $Nb_3Sn$ outsert has developed in Tohoku University, Sendai, Japan [40]. However, in the past years REBCO was gaining more and more interest and research activities on Bi-2223 were gradually reduced. Sumitomo is now the only manufacturer producing kilometer class long wires with an $I_c$ >200 A or wires of several hundred meters in length with $I_c$ >250 A [41].

In contrast to Bi-2223 tapes showing high anisotropy, Bi-2212 round wires are supposed to be promising in high-field applications for their isotropic property. As LTS wires, such round Bi-2212 wires can be easily made into insert coils for high-field NMR applications, Rutherford cables for the magnets of accelerators, or cable-in-conduit conductors for large magnets for fusion and detectors. At present, companies such as Showa (Japan), Oxford Superconducting Technologies (OST, USA) and Alcatel/Nexans (France) are able to produce kilometer class multifilamentary Bi-2212 long wires. In 2003, a Bi-2212 superconducting insert magnet generated a magnetic field of 5 T in a 20 T background field by Showa, which was used for a 950 MHz NMR system. In 2014, using OST Bi-2212 wires, the National High Magnetic Field Laboratory in the USA achieved a 34 T magnet that consist of a small Bi-2212 insert coil in a 31 T background field, demonstrating the potential of Bi-2212 superconductors for NMR system above 1 GHz [42]. Berkeley National Laboratory and Brookhaven National Laboratory in the USA have studied the construction of dipole magnet for accelerators using with Bi-2212 Rutherford cables. In China,

Institute of Plasma Physics, Chinese Academy of Sciences (IPPCAS) considered the Bi-2212 CIC conductor for the Chinese Fusion Experimental Tokamak Reactor (CFETR) [43]. In 2017, a sub-size, three stage rope-type cable containing 42 strands was manufactured by IPPCAS using Bi-2212 wires provided by Northwest Institute for Non-ferrous Metal Research, China [44].

2.2 REBCO

Compared with BSCCO superconductors, REBCO has much lower anisotropy and a much higher in-field $J_c$ at 77 K. However, the uniaxial (c-axis) aligned grain structure achieved in BSCCO wires by rolling or melting induced texture is unsatisfactory for REBCO to obtain a high inter-grain $J_c$ and a biaxial grain texture is necessary. In order to realize biaxial texture, other than the PIT fabricating route, REBCO films are deposited on biaxially textured buffer layers, which are coated on a long and flexible tape-like metal substrate. In addition, for environmental protection and thermal stabilization, an Ag layer a few μm thick and a thicker Cu layer are coated on the conductor. Since 2003, varieties of approaches for producing coated conductors have been implemented at industrial level [45]. The textured substrate techniques include ion beam assisted deposition (IBAD), rolling assisted biaxially textured substrates (RABiTS) and inclined substrate deposition. The deposition of epitaxial REBCO layer can be also achieved via various routes, including chemical routes, such as metal organic deposition (MOD) and metal organic chemical vapor deposition (MOCVD), or via physical routes such as pulsed laser deposition (PLD) and reactive co-evaporation (RCE). Among them, MOD and RCE are *ex situ* processes incorporating two steps: deposition of the pre-cursors and subsequent conversion of the precursors into REBCO. On the other hand, deposition and formation occur simultaneously during *in situ* processes such as PLD and MOCVD. The PLD process provides wide-ranging defect pinning structure, high-quality grain alignment and feasibility for thick-film deposition, but needs more expensive equipment and has a lower deposition rate. The MOCVD and RCE technique have high deposition rate and large deposition area, but their products have relatively poor inherent pinning microstructures and crystallinity. The MOD process has advantages such as low system cost and the least solution waste, but its products have larger pinning particles than that obtained in PLD growth, and the pinning at higher fields is less effective. The coated conductors manufactured with different routes can be used for different applications according to their performance [46, 47].

So far, long-length (>1000 m) REBCO coated conductors with critical current over 300 A/cm-width at 77 K in self-field have been achieved by several companies. In 2008, the world's first kilometer class REBCO tape was developed by SuperPower in the USA with IBAD/MOCVD process ($I_c$ = 227 A/cm at 77 K, in self-field) [48]. In 2016, SuNAM developed a ~1000 m long 12 mm wide REBCO tape with $I_c$ ~ 875 A/cm via IBAD/RCE process [49]. In the recent years, commercialization of REBCO coated conductors has been achieved by companies in the USA (ASMC, SuperPower, STI), Japan (SWCC, Fujikura, Sumitomo), Korea (SuNAM), Germany (THEVA, d-NANO, Bruke), Russia (SuperOx) and China (Shanghai Superconductor, Samri, SCSC), and rapid progress has been made for high-performance long-length REBCO coated conductors. The $I_c$ of their commercial coated conductors covers a wide range of values, from 100 A to over 250 A at 77 K, self-field (for a 4 mm wide tape) [45]. At present, the research for improving the current carrying ability for coated conductors focus on introducing artificial pinning centers (APCs) in the ceramic layer and increasing the thickness of the ceramic layer. The combination of the biaxial structure of the REBCO tapes with APCs such as non-superconducting nanoparticles or nanorods efficiently pins vortices at high temperatures, thus significantly enhances the in-field $J_c$ performance, as presented in figure 5. Moreover, it was observed that the APCs can be used to adjust the $J_c$ anisotropy, and are promising to further lower the anisotropy of REBCO tapes [45]. It was also found that the APCs can suppress the degradation of $J_c$ in REBCO tapes with the increased thickness of superconducting layers. With such improved flux pinning, ceramic layer with thickness of ~ 5 μm, which is 5 times thicker than that in commercial coated conductors, was achieved by SuNAM and KERI in Korea and University of Houston with a very high $I_c$ reaching 1300~1500 A/cm (77 K, self-field), respectively [51-53].

With the commercialization of REBCO coated conductors, lots of demonstration projects based on them for superconducting electric power devices such as power cables, motors, generators, transformers, and fault current limiters have been developed in Europe, USA, Japan, Korea and China [46, 48]. For example, in 2019 Korea Electric Power Corporation has fully funded and completed the first commercial project of HTS power cables in the real grid, called the Shingal Project, to connect two substations with a 23 kV HTS cable over a distance of 1 km [54]. Besides the power application projects there is continuous and remarkable progress of high-field applications using coated conductors, as shown in Figure 6. At present, the world record of DC magnetic field is 45.5 T, which was achieved by the National High Magnetic Field Laboratory (NHMFL, USA), with a 14.4 T REBCO insert in a 31.1 T resistive magnet [55]. For all-superconducting magnets, a 27.6 magnet with a Bi-2223 coil and a small REBCO test coil in a 17 T LTS background magnet was developed in RIKEN, Japan [56], a 32 T magnet with a 15 T LTS oursert and a 17 T REBCO insert was developed in NHMFL in 2017 [57], and in 2019 a 32.35 T magnet with a 15 T LTS oursert and a REBCO insert was developed in IEECAS [58]. A 26.7 T all-REBCO magnet was also developed by SuNAM [59]. These achievements in pursuing new records for high magnetic field clearly indicate the great potential of REBCO tapes in magnet applications.

## 3. MgB$_2$ superconductors

The MgB$_2$ superconductor discovered in early 2001 has a superconducting critical transition temperature as high as 39 K, setting a record for the $T_c$ of intermetallic superconducting materials. In contrast to cuprate HTS, the supercurrents in MgB$_2$ are not sensitive to the weak-linked grain boundaries [60], so MgB$_2$ is quite promising for fabricating high-performance wires. However, due to the weak flux pinning ability, the critical current density of MgB$_2$ drops rapidly with the increase of applied magnetic field, which limits the application of MgB$_2$ in high magnetic field region. Because of its relatively high $T_c$, low raw material cost, simple chemical composition and light weight, MgB$_2$ superconductor has also attracted wide attention from the applied superconductivity community. It is generally believed that $MgB_2$ superconducting materials have obvious technical and cost advantages in the application of superconducting magnets in MRI systems at 1-2 T and 10-20 K regions.

There are two main fabrication routes for $MgB_2$ wires: (1) PIT method: This process is relatively simple and has been widely used in the preparation of Bi-2223 and Bi-2212 wires and tapes. $MgB_2$ wires have been made both by *in situ* and *ex situ* PIT method. The *in situ* process starts by packing the powders of unreacted raw materials into metallic tubes, while the *ex situ* PIT method starts with reacted $MgB_2$ precursor powder packed into metallic tubes. Unlike Bi-2223 and Bi-2212 wires for which Ag or Ag alloy must be used as sheath materials, and $Nb_3Sn$ wires for which Cu alloys can be used as sheath materials, for $MgB_2$ wires since elemental diffusion and chemical reaction will occur between Ag and Mg or Cu and Mg, they usually prepared by using Fe, stainless steel and carbon steel sheaths, or composite sheaths composed of Cu alloys outer sheath and Nb, Ta, Fe inner barrier sheath. (2) Internal Mg diffusion process (IMD): this process involved an Mg rod placed in the center of the metal tube, with B powder filled between the metal tube and the Mg rod. After the assembly is cold deformed into wires, the Mg melts and diffuses into the surrounding B powder to form the $MgB_2$ superconducting phase during the final heat treatment. Compared with the commercialized PIT process, the IMD process is easy to obtain high-density $MgB_2$ phase, and thus achieving high transport $J_c$, so it has been the hotspot for the research of $MgB_2$ wires [61].

For $MgB_2$ superconductor, due to the lack of pinning centers inside the material, their current carrying ability was seriously suppressed by increasing field. The improvement of flux pinning is the key to enhance the $J_c$ performance of $MgB_2$ wires in strong magnetic fields. Currently carbon doping is the most effective and widely used method to enhance $J_c$. The doped carbon into $MgB_2$ lattice can substitute on the boron site, causing a decrease in the mean free path of electrons, which results in an increase of upper critical field. The most effective carbon dopents are recognized as nano-SiC and nano-C (or C-containing compounds) [62, 63]. With the carbon doping, the transport $J_c$ of MgB$_2$ wires based on PIT method can be significantly increased to $6\times10^4$ A/cm$^2$ at 4.2 K and 10 T, while the $J_c$ of wire samples by IMD method can be enhanced as high as $1.5\times10^5$ A/cm$^2$ [27].

At present, 100-meter-class MgB$_2$ wires based on IMD method have been achieved in several labs [61, 64], while companies such as Hyper Tech in the USA, Columbus in Italy, Hitachi in Japan, and Sam Dong in South Korea and Western Superconducting Technologies in China have possessed a production capacity of kilometer-level practical MgB$_2$ wires ($J_c$ = 1-2$\times10^5$ A/cm$^2$ at 4.2 K and 4 T) based on PIT method. Among them, the MgB$_2$ wires produced by Hyper Tech and Columbus have been successfully employed in applications such as magnetic resonance imaging, superconducting fault current limiter, and superconducting cables. In 2006, the Columbus developed the world's first MgB$_2$-based open MRI system. The system, which uses a GM refrigerator for cooling and can generate a magnetic field of 0.5-0.6 T at 20 K, obtained the first scan of the human brain [65]. So far, they have produced more than 20 sets of the above-mentioned MgB$_2$-based MRI system. It can be expected the operated field strength to be increasing to 1-2 T by using MgB$_2$ wires with higher $J_c$ performance in the future to satisfy the various medical diagnosis. Recently, the application of MgB$_2$ superconductor in the motors of wind power generation has received a significant boost. Hyper Tech has developed 8-20 MW wind power generators with superconducting stator and rotor based on low AC loss MgB$_2$ wires. Taking a 10 MW superconducting generator as an example, its weight is just about 50-60 tons, which is much lighter than the about 350 tons of a conventional generator.

## 4. Iron-based superconductors

Studies on the grain boundary nature in iron-based superconductors (IBS) suggested that intergrain currents across mismatched grains in iron-based superconductors are deteriorated to a lesser extent than in REBCO superconductors [66]. Therefore, the low-cost PIT method, which has been utilized in commercial Nb$_3$Sn、Bi-2223 and MgB$_2$ wires, is promising for IBS wires manufacture. At present, silver is widely used as sheath material for wires made of iron pnictides such as $Sr_{1-x}K_xFeAs$ (Sr-122) and $Ba_{1-x}K_xFeAs$ (Ba-122), since silver is chemically stable and not easy to react with iron pnictides during heat treatment of wires. On the other hand, in contrast to BSCCO wires, whose sheath material was limited to Ag or some Ag rich alloys due to the requirement for oxygen permeability, the IBS wires have more choices for sheath materials. For instance, a composite sheath consist of other cheap and stiff metals as the outer sheath and silver as the inner sheath can be used for IBS wires. The composite sheath can reduce the ratio of silver cost, provide sheath chemical stability, and enhance mechanical properties at the same time. Therefore, the low-cost, high-strength and high $J_c$ performance IBS wire and tape conductors are very promising based on PIT method.

In 2008, the first iron-based superconducting wires are developed in IEECAS by *in situ* PIT method, which starts by packing the powders of unreacted precursor materials into a metallic tube in a high purity Ar atmosphere. However, the defects in the material such as micro cracks, low density, phase inhomogeneity, and impurity phase restricted the transport current in wires. By using *ex situ* PIT method, in which reacted and well ground superconducting materials are packed into metallic tubes, the mass density and phase homogeneity of the wire after the final heat treatment are significantly improved in iron pnictide wires [67]. In the past years, mechanical deformation processes such as flat rolling, isostatic pressing and uniaxial pressing have significantly improved the $J_c$ of 122-type IBS in USA, Europe, China, Japan and Australia [68-72]. High transport $J_c$ above the practical level of $10^5$ A/cm$^2$ (4.2 K, 10 T) has been achieved by hot or cold uniaxial press combined with flat rolling process for PIT wires and tapes [73, 74]. A synergetic microstructural tailoring for mass density, grain alignment and micro-cracks is the key to realize such high $J_c$ performance. By using optimized hot press process to achieve a higher degree of grain texture, the transport $J_c$ was further increase to $1.5\times10^5$ A/cm$^2$ ($I_c$ = 437 A) with a small $J_c$ anisotropy of 1.37 at 4.2 K and 10 T in Ba-122 tapes, which exhibits very weak field dependence up to 33 T, as presented in Figure 7. The transport $J_c$ measured at 4.2 K under high magnetic fields of 27 T is still on the level of $5.5\times10^4$ A/cm$^2$, showing a great application potential in moderate temperature range which can be reached by liquid hydrogen or cryogenic cooling [75]. On the other hand, a high $J_c$ of $4\times10^4$ A/cm$^2$ was measured at 4.2 K and 10 T for Cu/Ag sheathed Ba-122 round wire processed with a hot isostatic pressing densification (HIP) [76], as shown in Figure 7. A local grain alignment perpendicular to the wire axis can be observed and ascribed to drawing or groove rolling process. Last year, a practical level critical current density up to $1.1\times10^5$ A cm$^{-2}$ at 4.2 K and 10 T was achieved in Cu/Ag sheathed Ba-122 tapes by combing flat rolling to induce grain texture and a subsequent HIP densification, which is a scalable and cost-effective manufacturing route [77].

For 11-type iron chalcogenides wires based on traditional PIT process, the most extensively used sheath material is Fe, since it has been proved to be the most chemically compatible with the 11-IBS phase. It also allows a PIT based diffusion process, where the Fe near the inner surface of the sheath and the Se and Te powder (for the *in situ* method) or FeTe$_{1-x}$Se$_x$ powder (for the *ex situ* method) inside the Fe tube form the superconducting phase by chemical reaction during the heat treatment process. However, the difficulty in controlling the Fe content and the decomposition of the Fe(Se,Te) phase during the heat treatment process restricts the transport $J_c$ of 11-IBS wires [73]. On the other hand, thin film technology exhibits its application potential for the fabrication of 11-IBS coated conductors. An almost isotropic $J_c$ of $1.7\times10^5$ A cm$^{-2}$, which is lowered by less than one order of magnitude in high fields up to 18 T, was achieved in Fe(Se,Te) film deposited on a RABiTS template [78]. Considering the simple elemental constituent, the moderate $T_c$ (~16 K), the high upper critical fields and the relative ease of fabrication for 11-IBS thin films, it is very promising for high-field applications at low temperature range [79]. Moreover, the newly discovered iron chalcogenides K$_x$Fe$_2$Se$_2$ [80] ($T_c$ > 30 K) and FeSe-based (Li,Fe)OHFeSe [81] ($T_c$ ~ 40 K) may broaden the application range of iron chalcogenides.

For practical applications of iron-based superconductors, fabricating wires and tapes with multifilaments in metal matrix to protect against flux jumps and thermal quenching is an important step. Based on the techniques used in the single-core IBS wires, Fe/Ag clad 7-filament 122-IBS wires and tapes were successfully fabricated with the PIT process by IEECAS in 2013 [82]. After that, Fe/Ag sheathed 114-filament 122-IBS wires and tapes were also produced [29]. Processed with hot press, a high transport $J_c$ of $3.6\times10^4$ A/cm$^2$ (4.2 K, 10 T) can be achieved in 7-filament Monel/Ag sheathed 122-IBS tapes, which exhibit an improved mechanical strength and very weak field dependence for transport $J_c$ [74]. Though high $J_c$ properties can be obtained in short '122' IBS samples, practical applications need wire and tape conductors with sufficient length. In 2014, the IEECAS group fabricated the first 11 m long 122-IBS tape by a scalable rolling process. After carefully optimizing the long-length wire fabricating process to achieve a higher-level uniformity of deformation, the world's first 100 meter-class IBS tapes was produced by the same group [83]. This 115 m long 7-filament Sr-122 tape shows a uniform $J_c$ distribution throughout the tape with a minimum $J_c$ of $1.2\times10^4$ A/cm$^2$ (4.2 K, 10 T). Very recently, by improving the fabrication process, a 100-m 7-filament Ba-122 tape with $J_c$ above $5.0\times10^4$ A/cm$^2$ (4.2 K, 10 T) was achieved, demonstrating great potential in large-scale manufacture and a promising future of iron-based superconductors for practical applications.

In 2018, an IBS single pancake coil was firstly fabricated with Ba-122 superconducting tapes and tested under a 24 T background field [84], then another single pancake coil and a double pancake coil were developed and tested at a 30 T background field [85], showing very weak dependence of critical currents on such high magnetic field. In 2020, by using 100-meter 7-filament Ba-122 tapes provided by IEECAS, IBS racetrack coils were firstly fabricated at Institute of High Energy Physics, Chinese Academy of Sciences (IHEPCAS). The racetrack coils were tested in a low-temperature superconducting common-coil dipole magnet which provided a maximum background field of 10 T at 4.2 K. One of the best IBS racetrack coil quenched at 4.2 K and 10 T showed an operating current of 65 A, which is still as high as 86.7% of the $I_c$ of short samples at 10 T [86]. As presented in Figure 8, these results demonstrate that the IBS conductor is a promising candidate for the application of high field magnets especially for future high-energy accelerators. At present a conceptual design study of 12 T dipole magnets is ongoing with the IBS technology, to fulfill the requirements and need of a large-scale superconducting accelerator proposed by IHEPCAS [87].

## Future perspectives of materials and their applications

Nowadays, the inner force that promoting the development of superconducting materials with higher performance is mainly the great demand from high-field magnet applications such as MRI, NMR and large scientific projects including accelerators and fusion, since they greatly challenge for the performance limit such as transport critical density, upper critical field, mechanical strength, magnetic and thermal stability. The generation of magnetic fields exceeding the limits of LTS conductors provides considerable opportunities for HTS materials, and will speed up their commercialization. Compared to HTS materials which can be used in higher temperatures and magnetic fields, the advantages of LTS materials are cost, technology maturity and batch stability, which make them still the priority choice if the required temperature and field strength are within their performance limit. Moreover, when today's target magnetic fields of new superconducting magnets are higher than the working limit of LTS wires and HTS wires are indispensable, LTS magnets are still used to provide background fields for central HTS magnets to save the cost of the whole superconducting magnets.

For commercial applications with superconducting magnets, MRI and NMR systems are growing exponentially and fostered a corresponding exponential growth in the wire production capacity. In 2019, the Iseult 11.7 T whole body MRI system has been constructed by CEA in France, and for the next goal will be a >14 T whole body MRI system, for which the superconducting wires used should be $Nb_3Sn$ or HTS instead of the current Nb-Ti [88]. The development of biology and novel drugs is calling for >1-1.3 GHz NMR system, which need 25-30 T superconducting magnets that using HTS wires are indispensable. On the other hand, large scientific projects such as the Large Hadron Collider at CERN (built between 2002 and 2008) and ITER have also significantly benefited the superconducting material industry. Some large projects ahead are the Future Circular Collider (FCC) at CERN and the large tokamaks (such as EU-DEMO in Europe, K-DEMO in Korea, and CFETR in China). While a Circular Electron Positron Collider (CEPC) and its upgraded version Super Proton Proton Collider (SPPC) are under consideration in China [87]. The accelerator magnets for FCC and SPPC are required to generate 16-20 T fields, for which HTS conductors are preferred.

On the other hand, superconductivity applications for electric power technologies are still exploring their niche in the market, some application scenario such as superconducting electric power cables and superconducting maglev trains for big cities, superconducting power station connected to renewable energy network, and liquid hydrogen or LNG cooled electric power generation/transmission/storage system at ports or power plants may achieve commercialization in the future. In Japan, the superconducting maglev test track in Yamanashi, which has set a new speed record of 603 km/h of rail vehicles in 2015, is planned to be expanded into a commercial line linking Tokyo and Nagoya in 2027. In the USA and Europe, NASA and Airbus have started their own development project for electric aircraft. With the power distributed electrically from turbine-driven superconducting generators to superconducting motors that drive electric fans for propulsion, the E-aircraft have high fuel efficiency, less engine noise, and can contribute to the reduction of greenhouse gas emissions. Due to the limit capability of cooling system on aircraft, the HTS power and transmission technology will be used. In China, started in 2020, two commercial demonstration projects of REBCO power cables cooled with liquid nitrogen are under construction in the downtown of Shanghai and Shenzhen by State Grid Corporation of China and China Southern Power Grid, respectively.

For the above mentioned large machines and big projects, the cost of superconductors becomes a crucial issue. For cuprate superconductors that are stepping into commercialization, the product price is still the main obstacle for their large-scale application. The current price is about 5 $/kA m for $Nb_3Sn$, 60-80 $/kA m for Bi-2212 and Bi-2223 and 100-200 $/kA m for REBCO conductors for use at 4.2 K and 10 T [34].Their price is still higher than the ideal 25 $/kA m for large-scale applications. Therefore, lowering the price of is the top priority for their research and development. For REBCO coated conductors, one of the reasons for their high price seems to be the low manufacturing yield of long-length tape products, especially for the ones longer than 500 m [34]. Another factor that raises the price is the quite low ratio (just a few percent of the whole cross section of the tapes, much lower than the 20-40% for BSCCO PIT wires and tapes) of the superconducting layer. Therefore, increasing the thickness of the superconducting layer is another important issue for the low-cost produce for REBCO tapes.

For Bi-2212 and Bi-2223 wires and tapes, besides the price, their relatively weak mechanical strength due to the presence of soft Ag or Ag alloys sheath is another critical issue that should be taken into consideration for applications under high electromagnetic stress. The weak mechanical strength of wires and tapes usually require a wind-and-react method for magnet constructions. However, the reaction by heat treatment can largely degrade the mechanical strength of the structural materials of magnets, which have become one of the main obstacles for ultra-high magnets with field strength close to 40 T. Laminated mechanical reinforcement technique has been successfully employed by Sumitomo for commercial Bi-2223 tapes, but they are still facing the competition with the fast-growing REBCO tapes. On the other hand, Bi-2212 round wires can still find a market for high-field applications because of their isotropic property and flexible architecture. Bi-2212 round wires can be inserted in high strength alloy tubes for reinforcement, and there have been several studies on the structural material for the wind-and-react process for Bi-2212 wires [89, 90].

For $MgB_2$ and iron-based superconductors, as mentioned in the above sections, metallic composite sheaths can be used for wire fabrication. Therefore, their mechanical strength would not be the main drawback for their practical applications. For PIT $MgB_2$ wires, the critical current density in fields is still needed to improve to make them to be better used at around 20 K; for IMD $MgB_2$ wires, the proportion of superconducting layer is needed to increase as for REBCO tapes. For iron-based superconducting wires, it seems that grain texture is crucial for obtaining superior $J_c$, since now the best $J_c$ in textured tapes is about three times higher than that in wires.

The composite sheaths for iron-based superconducting wires contain an inner Ag barrier sheath, whose melting temperature (lower than that of IBS phase) restricts the use of melting texture process for round wires as used for Bi-2212. Nevertheless, for the tape conductors, it is possible to orient the tapes properly in the preferred direction to the magnetic field in high-field applications, so as to make use of their best transport and mechanical properties. Moreover, the relatively smaller critical bending radius of tapes would enable the react-and-wind process other than the wind-and-react process, thus expanding the choice of structural and insulating materials for the construction of magnets [34]. In addition to their low intrinsic anisotropy, iron-based superconductors can still become an ideal candidate for high-field applications after $J_c$ being further increased in the future.

## Conclusion

At present the great demand from high-field magnet applications provides a technology pull for HTS conductors, and will speed up their commercialization. On the other hand, superconductivity applications for electric power technologies are still exploring their niche in the market. Compared with HTS materials which can be used in higher temperatures and magnetic fields, the advantages of LTS materials are cost, technology maturity and batch stability, which make them still the priority choice if the required temperature and field strength are within their performance limit. The main obstacle for the large-scale application of BSCCO and REBCO conductors is still the high price, so it is urgent to increase their performance/cost ratio. The feasibility of metallic composite sheaths are advantageous for the high-mechanical strength and low-cost fabrication for $MgB_2$ and iron-based superconductors, but the current carrying capability still need to be enhanced before gaining large-scale applications.

## Acknowledgments

This work is supported by the National Key R&D Program of China (Grant Nos. 2018YFA0704200 and 2017YFE0129500), the National Natural Science Foundation of China (Grant Nos. 51861135311, U1832213 and 51721005), the Strategic Priority Research Program of Chinese Academy of Sciences (Grant No. XDB25000000), and the Key Research Program of Frontier Sciences of Chinese Academy of Sciences (Grant No. QYZDJ-SSW-JSC026).

## Author contributions

Writing - original draft, C.Y.; writing - review & editing, C.Y., and Y.M.; supervision, Y.M.

## Declaration of interests

The authors declare no competing interests.

## Captions

Figure 1. Main applications of superconducting power and magnetic technologies with their typical operating magnetic fields. The areas of each application are marked by coloured frames.

Figure 2. Evolution of high temperature superconductivity over time. The boiling temperatures of coolants such as liquid nitrogen (LN$_2$) and liquid hydrogen (LH$_2$) are marked by dashed lines.

Figure 3. Comparative $H$-$T$ phase diagram for representative cuprates, iron-based superconductors, and conventional superconductors, where the solid and dashed lines show, respectively, upper critical field $H_{c2}(T)$ and irreversibility field $H^*(T)$ parallel to the c-axis. For anisotropic superconductors, the $J_c(H)$ vanishes at $H^*(T)$, which can be much smaller than $H_{c2}(T)$. The data are collected from [17]. (LHe: liquid helium; LN$_2$: liquid nitrogen, LH$_2$: liquid hydrogen).

Figure 4. Typical images of the cross sections of wires and tapes and the development status for practical superconducting materials including niobium-based superconductors (Nb-Ti and Nb$_3$Sn) [23], copper-based oxides (Bi-2223 [24], Bi-2212 [25], REBCO [26, 23], MgB$_2$ [27, 28] and iron-based superconductors (IBS) [29].

Figure 5. Enhancement of the critical current density $J_c(B)$ associated with the two big boosts of the coated conductors development: (1) development of the biaxial texture onto metallic substrates and (2) introduction of the nanocomposite artificial pinning centers (APCs) in the coated conductors. With the combination of biaxial texture structure and nanocomposite APCs, the $J_c$ of coated conductors is much higher than that of Nb-Ti and Nb$_3$Sn wires. The data are collected from [46, 50].

Figure 6. State-of-the-art high-field solenoid magnets based on HTS materials: (a) a 45.5 T hybrid magnet with a 14.4 T REBCO insert in a 31.1 T resistive magnet [55] and a 32 T magnet with a 15 T LTS oursert and a 17 T REBCO insert [57]; (b) a 32.35 T magnet with a 15 T LTS oursert and a REBCO insert [58]; (c) a 26.7 T all-REBCO magnet [59]; (d) a 27.6 T magnet with a Bi-2223 coil and a small REBCO test coil in a 17 T LTS background magnet [56]; (e) a 24.6 T cryogen free magnet with a Bi-2223

insert and an Nb$_3$Sn outsert [40].

Figure 7. State-of-the-art critical current density $J_c$ of 122-IBS based on PIT technique showing much weaker field dependence up to 33 T compared with that of Nb-Ti and Nb$_3$Sn LTS conductors. Data source for IBS short samples: Ba-122 tapes ($J_c$ at 10-13 T, 24-27 T and 25-33 T were measured in a 14 T superconducting magnet, a 28 T hybrid magnet and a 35 T water-cooled magnet, respectively) [74, 75]; Ba-122 wires [76]. Data for LTS conductors were collected from nationalmaglab.org [23].

Figure 8. Pancake coils [84, 85] and racetrack coils [86] made with iron-based superconductors and their in-field critical currents, which show weak field dependence up to 30 T and 12 T, respectively.

## Table

Table 1 Basic material and critical current density relevant parameters for practical superconductors and their wire fabrication technology and typical forms at present

| Material | $T_{c, max}$ (K) | $H_{c2, 4.2 K}$ (T) | $J_{c, 4.2k}$ (A/cm$^2$) | Coherence length $\varepsilon_{ab}$ (nm) | Anisotropy $\gamma_H$ | Wire technology | Typical wire forms |
|---|---|---|---|---|---|---|---|
| Nb-Ti | 9.5 | 11.5 | 4×10$^5$ (5 T) | 4 | Negligible | \ | round wire |
| Nb$_3$Sn | 18 | 25 | ~10$^6$ | 3 | Negligible | bronze process / internal Sn process / powder-in-tube | round wire |
| MgB$_2$ | 39 | 18 | ~10$^6$ | 6.5 | 2~2.7 | powder-in-tube / internal Mg diffusion | wire or tape |
| REBCO | 92 | >100 | ~10$^7$ | 1.5 | 5~7 | coated conductor | flat tape |
| Bi-2223 | 108 | >100 | ~10$^6$ | 1.5 | 50~90 | powder-in-tube | flat tape |
| Bi-2212 | 90 | | | | | powder-in-tube | wire or tape |
| 1111 IBS | 55 | >100 | ~10$^6$ | 1.8~2.3 | 4~5 | powder-in-tube | wire or tape |
| 122 IBS | 38 | >80 | ~10$^6$ | 1.5~2.4 | 1.5~2 | powder-in-tube | wire or tape |
| 11 IBS | 16 | >40 | ~10$^5$ | 1.2 | 1.1~1.9 | powder-in-tube | wire or tape |

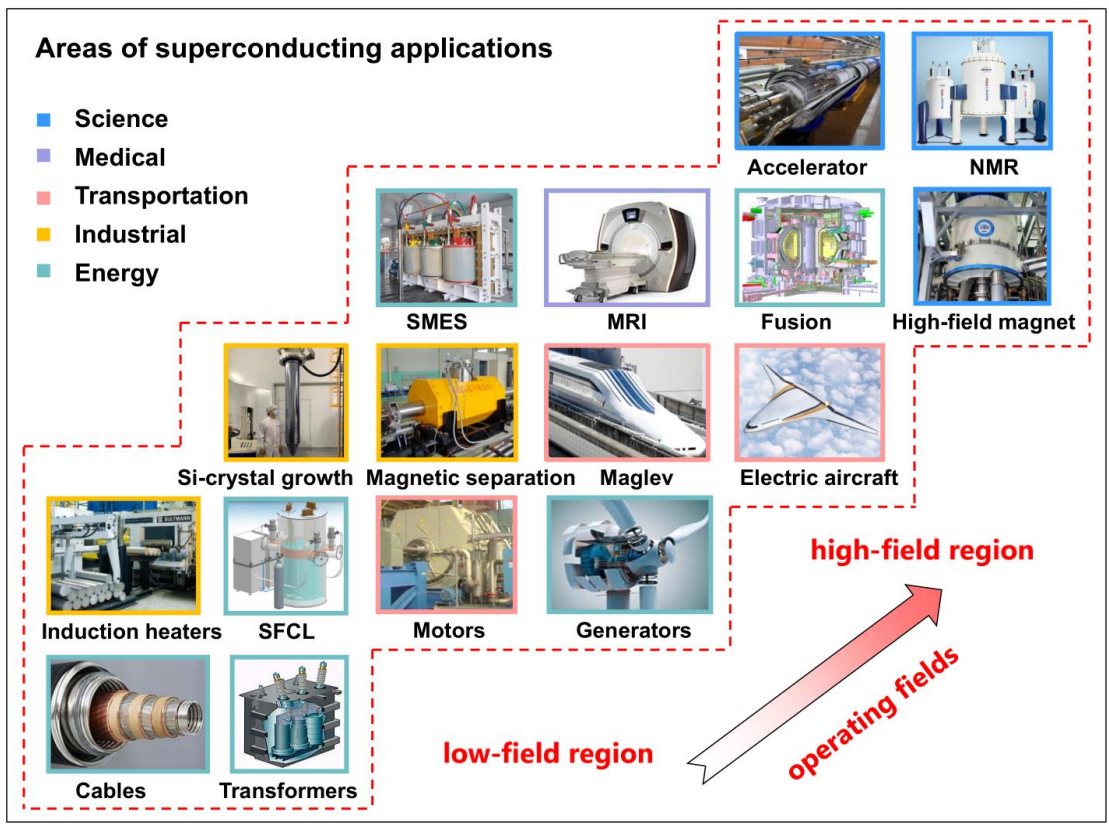

Figure 1.

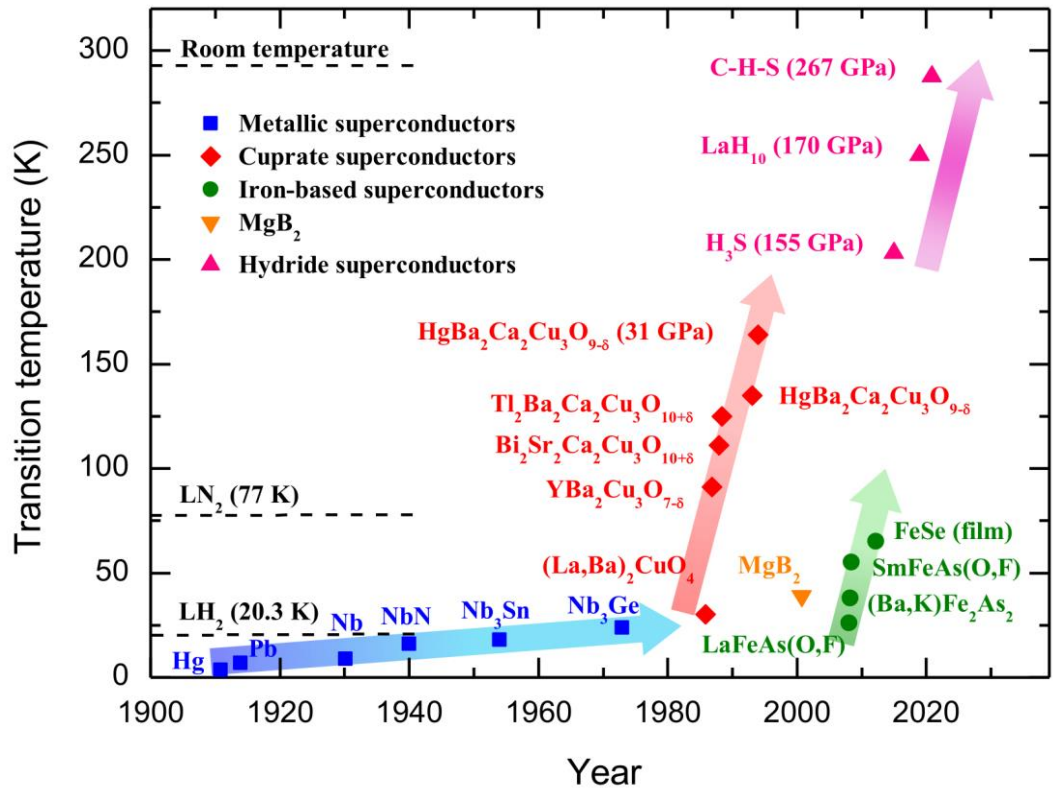

Figure 2.

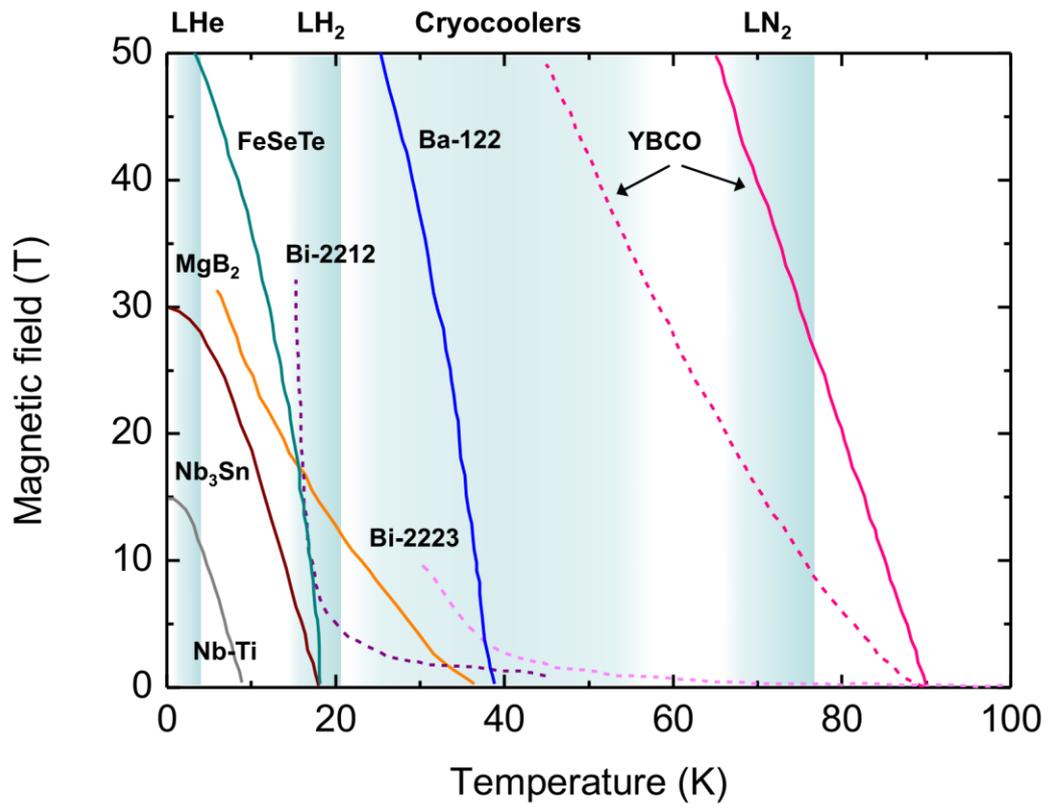

Figure 3.

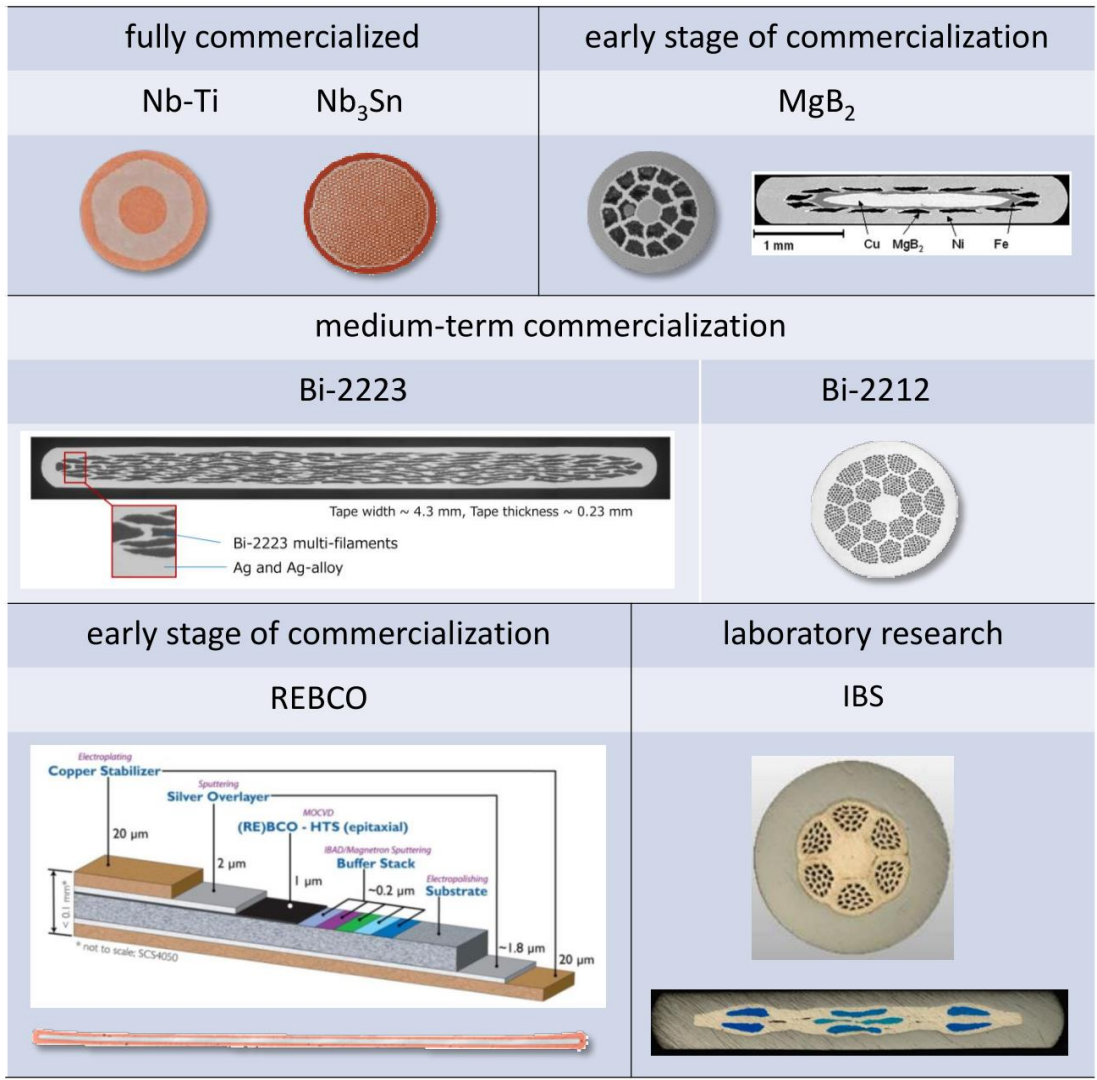

Figure 4.

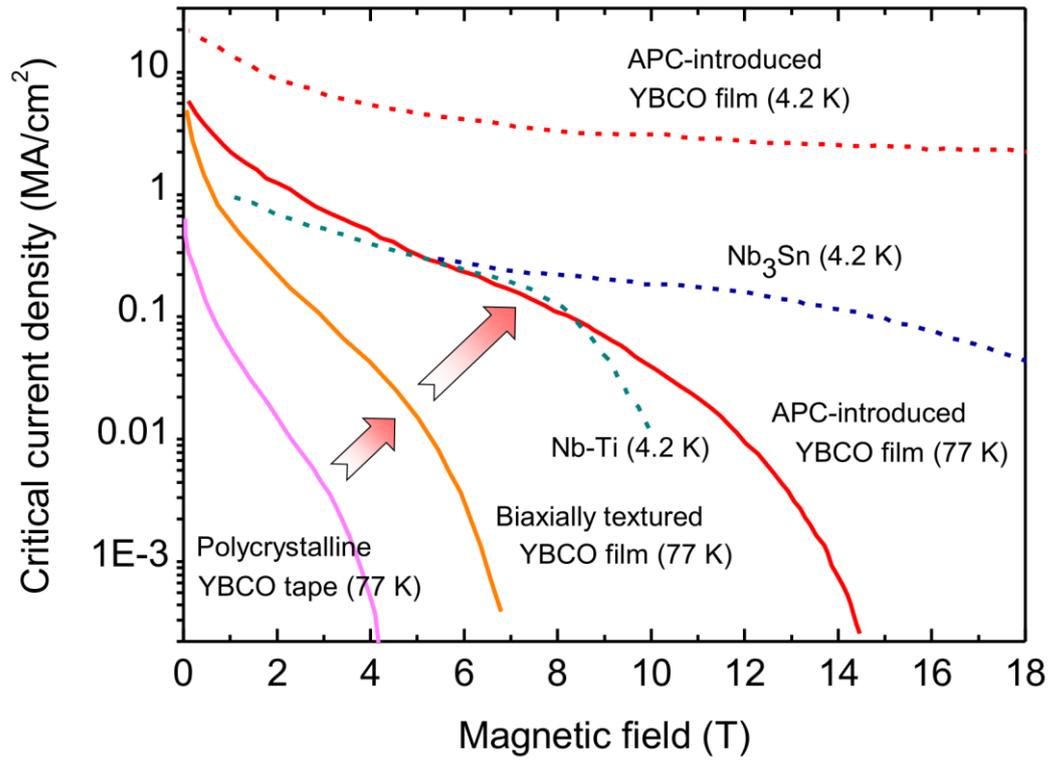

Figure 5.

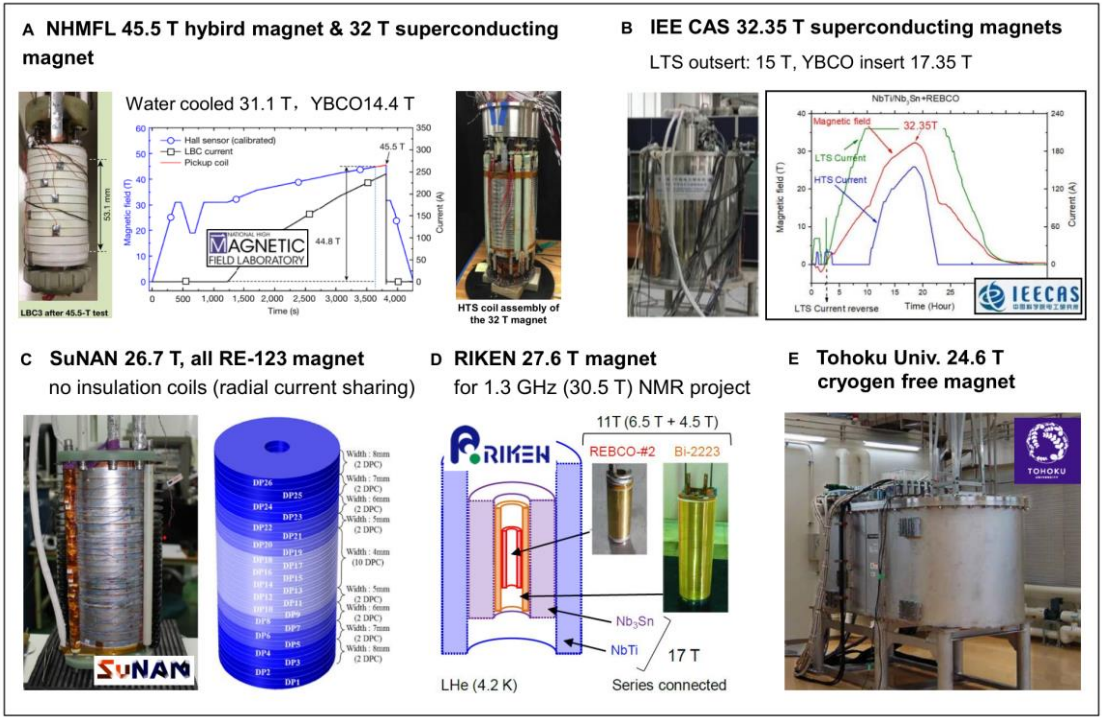

Figure 6.

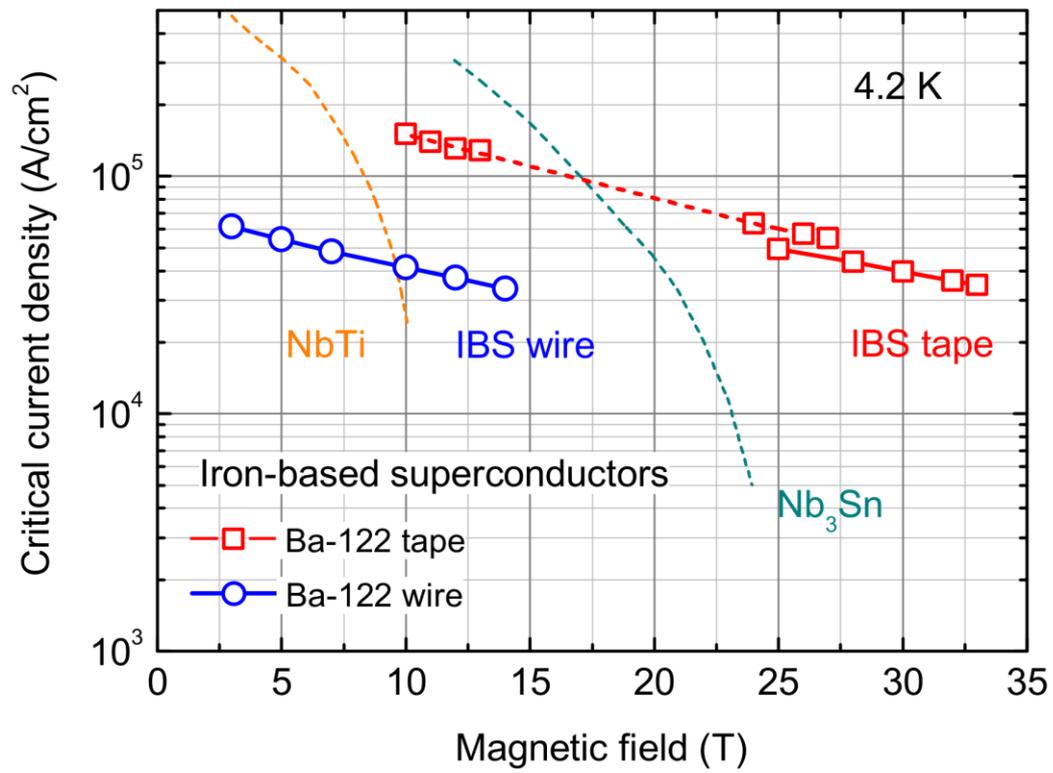

Figure 7.

Figure 8.